\documentclass[aps,prl,twocolumn,superscriptaddress]{revtex4-1}

\usepackage{graphicx} 
\usepackage{dcolumn}
\usepackage{bm}
\usepackage{amssymb,amsmath}
\usepackage{epstopdf}

\begin{document}

\title{Persistent Optically Induced Magnetism in Oxygen-Deficient Strontium Titanate}

\normalsize
\author{W.~D.~Rice}
\affiliation{National High Magnetic Field Laboratory, Los Alamos National Laboratory, Los Alamos, NM 87545, USA}
\author{P.~Ambwani}
\affiliation{Department of Chemical Engineering and Materials Science,
University of Minnesota, Minneapolis, MN 55455, USA}
\author{M.~Bombeck}
\affiliation{Experimentelle Physik 2, Technische
Universit$\ddot{a}$t Dortmund, D-44221 Dortmund, Germany}

\author{J.~D.~Thompson}
\affiliation{Materials Physics and Applications, Los Alamos National
Laboratory, Los Alamos, NM 87545, USA}

\author{C.~Leighton}
\affiliation{Department of Chemical Engineering and Materials Science,
University of Minnesota, Minneapolis, MN 55455, USA}

\author{S.~A.~Crooker}
\affiliation{National High Magnetic Field Laboratory, Los Alamos National Laboratory, Los Alamos, NM 87545, USA}
\date{\today}


\maketitle


\textbf{Strontium titanate (SrTiO$_3$) is a foundational material in the emerging field of complex oxide electronics~\cite{RameshMRSBulletin2008, CenScience2009, MannhartScience2010, ChambersAdvMat2010, ZubkoAnnRevCondPhys2011, HwangNatureMat2012}.  While its electronic and optical properties have been studied for decades~\cite{MullerPRB1979, LeePRB1975, FaughnanPRB1971, WildPRB1973}, SrTiO$_3$ has recently become a renewed materials research focus catalyzed in part by the discovery of magnetism and superconductivity at interfaces between SrTiO$_3$ and other oxides~\cite{HwangNatureMat2012, BrinkmanNatureMat2007, DikinPRL2011, LiNaturePhys2011, BertNaturePhys2011, AriandoNatureComm2011, MoetakefPRX2012}. The formation and distribution of oxygen vacancies may play an essential but as-yet-incompletely understood role in these effects~\cite{MullerNature2004, EcksteinNatureMat2007, KalabukhovPRB2007, ShenPRB2012, PavlenkoPRB2012}. Moreover, recent signatures of magnetization in gated SrTiO$_3$~\cite{LeePRL2011, LeeGoldmanPRL2011} have further galvanized interest in the emergent properties of this nominally nonmagnetic material. Here we observe an \emph{optically induced} and \emph{persistent} magnetization in oxygen-deficient SrTiO$_{3-\delta}$ using magnetic circular dichroism (MCD) spectroscopy and SQUID magnetometry. This zero-field magnetization appears below $\sim$18K, persists for hours below 10K, and is tunable via the polarization and wavelength of sub-bandgap (400-500nm) light. These effects occur only in oxygen-deficient samples, revealing the detailed interplay between magnetism, lattice defects, and light in an archetypal oxide material.}

\begin{figure*} [htbp]
\centering
\includegraphics [scale = 0.75] {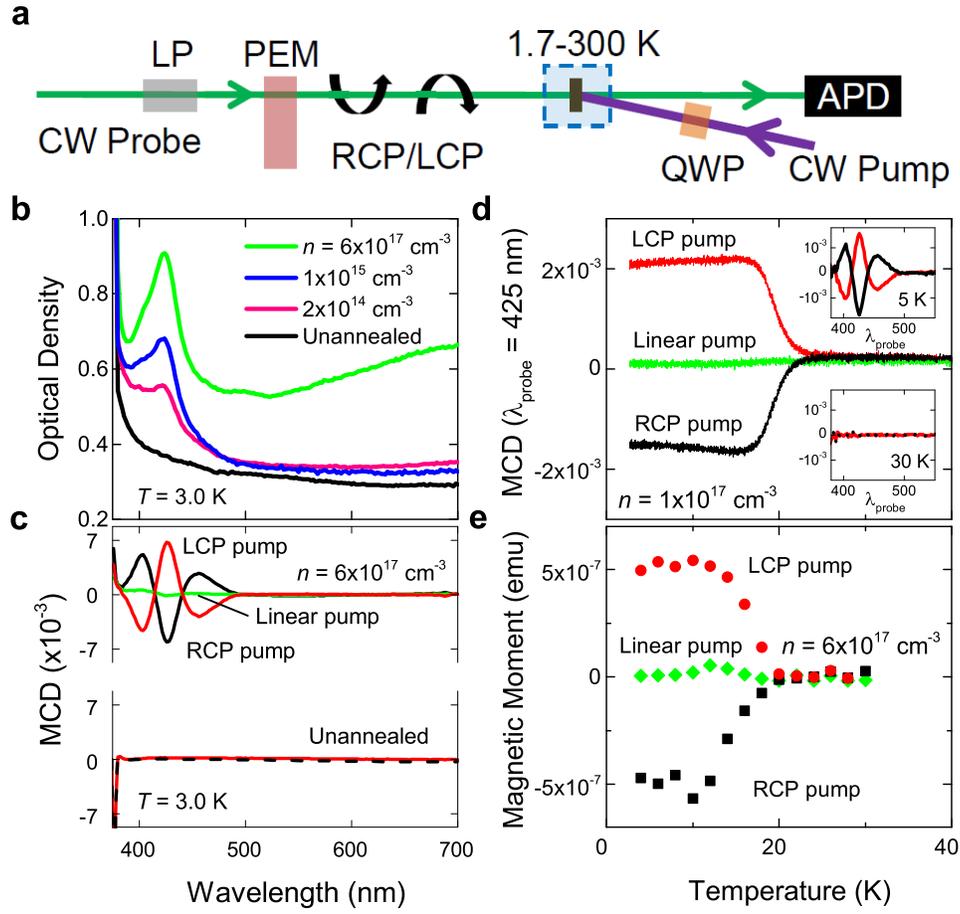}
\caption{\fontsize{10pt}{10pt}\selectfont
\textbf{Optically induced magnetization in oxygen-deficient SrTiO$_{3-\delta}$ at zero applied magnetic field.~a,}  Schematic of the magnetic circular dichroism (MCD) experiment used to detect magnetization in SrTiO$_{3-\delta}$ single crystals. Continuous-wave (CW) probe light is modulated between right- and left-circular polarization (RCP/LCP) by a linear polarizer (LP) and photoelastic modulator (PEM), then transmitted through the samples and detected by an avalanche photodiode (APD). The polarization of the additional CW pump light is controlled with a quarter-wave plate (QWP).~\textbf{b,}~Optical absorption spectra of several SrTiO$_{3-\delta}$ crystals at low temperature (3~K) and at zero magnetic field. The samples are denoted by their electron density, $n$, from which the $V_{\rm O}$ concentration may be inferred. Optical density=-ln(T/T$_0$ ); these spectra have not been corrected for simple Fresnel reflection.~\textbf{c,} The corresponding MCD spectra from a SrTiO$_{3-\delta}$ and an as-received (unannealed) SrTiO$_3$ single crystal after being weakly illuminated with 50~$\mu$W of RCP, LCP, and linearly polarized pump light at 405~nm (black, red, and green curves, respectively). The oscillatory MCD signals indicate a pump-induced magnetization, which inverts sign upon switching between RCP and LCP illumination.~\textbf{d,}~The temperature dependence of the induced magnetization (as monitored by MCD at $\lambda_{\rm probe}$=425~nm), while being pumped at 405~nm. Insets: MCD at 5~K and 30~K.~\textbf{e,}~The temperature dependence of the optically induced magnetization as measured by SQUID magnetometry, showing similar behavior as in~\textbf{d}.}
\label{Figure1}
\end{figure*}

To explore the relationship between oxygen vacancies ($V_{\rm O}$) and magnetism in SrTiO$_3$, we prepared a series of slightly oxygen-deficient SrTiO$_{3-\delta}$ single crystal samples by annealing (\emph{i.e.}, reducing) commercial SrTiO$_3$ substrates in ultra-high vacuum at varying temperatures. Isolated $V_{\rm O}$ in SrTiO$_3$ are shallow donors: if every $V_{\rm O}$ donates one to two electrons to the conduction band, the total $V_{\rm O}$ concentration can be approximately inferred by measuring the electron density $n$~\cite{SpinelliPRB2010}. Here, $n$ ranged from $\sim$$3\times 10^{12}$~cm$^{-3}$ to $8 \times 10^{17}$~cm$^{-3}$. To probe magnetism in these samples we use MCD spectroscopy, wherein small differences between the transmission of right- and left-circularly polarized (RCP/LCP) probe light are sensitively measured (Fig.~\ref{Figure1}a; see Methods). Non-zero MCD signals typically imply the presence of broken time-reversal symmetry (\emph{e.g.}, magnetization)~\cite{StephensJChemPhys1970}. The samples could also be weakly illuminated with a separate source of wavelength-tunable, polarization-controlled pump light.

Figure~\ref{Figure1}b displays optical absorption spectra from several SrTiO$_{3-\delta}$ single crystals. As-received (unannealed and nominally undoped) substrates show only the sharp onset of band-edge absorption at 380 nm (3.26~eV); at longer wavelengths the absorption is small, with a weak sub-bandgap tail. In contrast, increasingly oxygen-deficient SrTiO$_{3-\delta}$ crystals develop an additional sub-bandgap absorption peaked at $\sim$430~nm (380~meV below the band-edge), with a weak shoulder at $\sim$400~nm. However, this absorption does not scale linearly with $n$, strongly suggesting that only a fraction of the total $V_{\rm O}$ density contributes to this absorption peak, for example via specific $V_{\rm O}$-related complexes or clusters ~\cite{MullerNature2004}. Early studies of bulk SrTiO$_3$ revealed similar absorption peaks in this spectral range that were associated with Fe dopants and Fe-$V_{\rm O}$ complexes ~\cite{FaughnanPRB1971, WildPRB1973, LeePRB1975}.  Iron is only one of many metal impurities found in significant quantities even in nominally ``pure" SrTiO$_3$ crystals~\cite{SonNatMat2010} and thus an unambiguous assignment is difficult~\cite{LeePRB1975}.

Surprisingly, Fig.~\ref{Figure1}c shows that these $V_{\rm O}$-related absorption features are accompanied by the ability to induce -- and control -- a robust MCD signal at zero magnetic field by illuminating with weak, circularly polarized pump light at sub-bandgap wavelengths (here, $\lambda_{\rm pump}$=405~nm, which pumps the entire sample thickness). The optically induced MCD signal oscillates with probe wavelength over the same spectral range where the $V_{\rm O}$-related absorption occurs, exhibiting peak amplitudes at $\sim$400, 425, and 455~nm.  Moreover, the MCD signal exactly inverts when the pump polarization is switched from RCP to LCP and disappears when the pump is linearly polarized. In contrast, unannealed SrTiO$_3$ exhibits no optically induced MCD. Importantly, \emph{all} SrTiO$_{3-\delta}$ crystals showing a measurable sub-bandgap absorption peak (those with $n$$\gtrsim$10$^{14}$~cm$^{-3}$) demonstrate optically induced MCD with identical spectral shape, showing peaks and nodes at the \emph{same} $\lambda_{\rm probe}$. These data point to an optically induced magnetization arising from localized $V_{\rm O}$-related complexes.

The temperature dependence of the optically induced magnetization is shown in Fig.~\ref{Figure1}d. To probe magnetization continuously, we monitor the MCD signal at its peak ($\lambda_{\rm probe}$=425~nm).  Under steady optical pumping, the induced magnetization is approximately constant at low temperatures $T$, but abruptly disappears for $T$$\gtrsim$18~K. All of the SrTiO$_{3-\delta}$ samples display the same temperature dependence regardless of $n$, which again suggests localized independent complexes, rather than collective phenomena such as ferromagnetism.  The 430~nm absorption does not significantly change above 18~K (Supplementary Fig.~S1).

Conventional SQUID magnetometry was used to directly confirm and quantify the induced magnetization (see Methods).  Here, polarization-controlled 405~nm pump light was coupled to the samples via optical fiber.  Figure~\ref{Figure1}e shows an induced magnetic moment of $\sim$5$\times 10^{-7}$~emu at zero applied field, following the same temperature and polarization dependence as measured by MCD.  A signal of this magnitude is expected in this sample if every $V_{\rm O}$ induces a moment of $\sim$0.01~$\mu_{\rm B}$. More likely, only a small fraction of the $V_{\rm O}$ may induce a (much larger) moment, a scenario consistent with only a subset of the $V_{\rm O}$ density contributing to optically induced magnetization.  As with the MCD data, no SQUID signal was observed in unannealed samples (Fig.~S2).

\begin{figure*} [htbp]
\centering
\includegraphics [scale = 0.75] {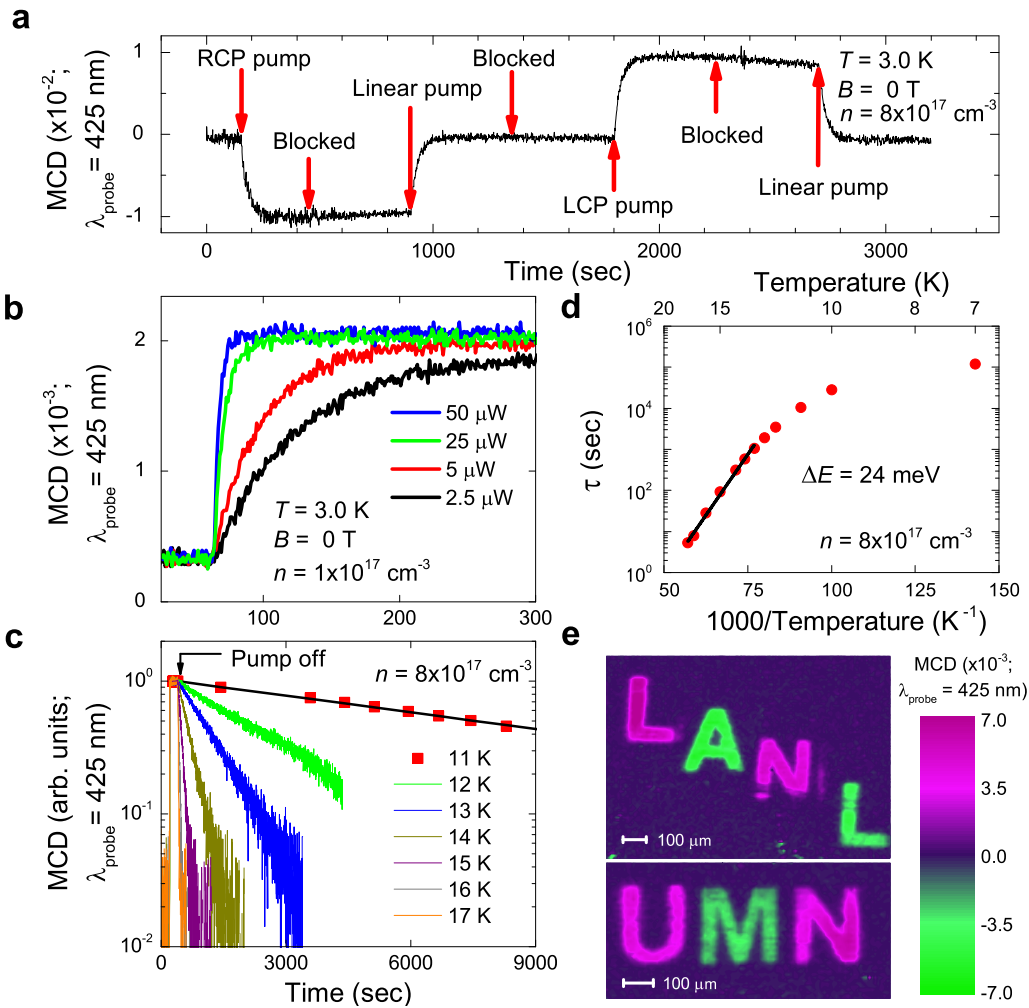} 
\caption{\fontsize{10pt}{10pt}\selectfont
\textbf{Persistence of the optically induced magnetization in SrTiO$_{3-\delta}$.~a,}~Temporal evolution of the magnetization (as monitored by MCD at $\lambda_{\rm probe}$=425~nm) for different polarizations of 405~nm pump light.  The optically induced magnetization persists even after the pump is blocked, and the pump polarization can control and invert the magnetization.~\textbf{b,}~As the pump intensity increases, the induced magnetization saturates more quickly.~\textbf{c,}~ Measuring the slow exponential decay of the optically induced magnetization at different temperatures. Relaxation at 11~K was probed intermittently to limit any probe-induced relaxation.~\textbf{d,}~$\tau$ increases over four orders of magnitude as temperature is reduced from 17~K to 7~K. An Arrenhius fit (black line) to the linear portion of the data gives an activation energy $\Delta E$=24~meV.~\textbf{e,}~A demonstration that magnetic information can be optically written into SrTiO$_{3-\delta}$, stored, and then optically read out.}
\label{Figure2}
\end{figure*}

Remarkably, the optically induced magnetization in SrTiO$_{3-\delta}$ is extremely long-lived at low temperatures,~\emph{i.e.} the magnetization persists long after the pump illumination is turned off. Figure~\ref{Figure2}a shows the induced magnetization over $\sim$1 hour as the pump illumination is varied. The magnetization is initially zero. At $t$=150~s, RCP pump light illuminates the sample, causing the magnetization to build up and saturate within about one minute. Unexpectedly, this magnetization \emph{persists} after the pump light is blocked at $t$=450~s. Subsequent pumping with linearly polarized light causes the magnetization to rapidly re-equilibrate back to zero. Using LCP pump light, equivalent but oppositely oriented magnetization dynamics are produced.  Similar temporal behavior is observed using SQUID magnetometry (Fig.~S2).

Figure~\ref{Figure2}b shows that the induced magnetization grows more rapidly with increasing pump intensity, but saturates at approximately the same value.  Despite achieving saturation, the induced MCD signal is only $\sim$10$^{-3}$, indicating that the total absorption at this wavelength ($\lambda_{\rm probe}$=425~nm) changes only minimally (see also Fig. S3).

The relaxation rate of the induced magnetization after the pump light is blocked is strongly temperature dependent. The magnetization relaxation fits very well to a single-exponential decay with time constant $\tau$ (Fig.~\ref{Figure2}c). Fig.~\ref{Figure2}d shows that at 17~K relaxation occurs within seconds; however, as the temperature drops, $\tau$ increases to several \emph{hours} (see also Fig.~S4). Data between 13 and 17~K suggest an approximately activated (Arrhenius) behavior $\left[\tau = \tau_0 \exp\left(\Delta E/k_{\rm B}T\right)\right]$, from which an activation energy $\Delta E$=24~meV is inferred. Below 13~K, $\tau$ deviates from activated behavior, but this may be due to small amounts of unintended light on the samples.

\begin{figure*} [htbp]
\centering
\includegraphics [scale = .75] {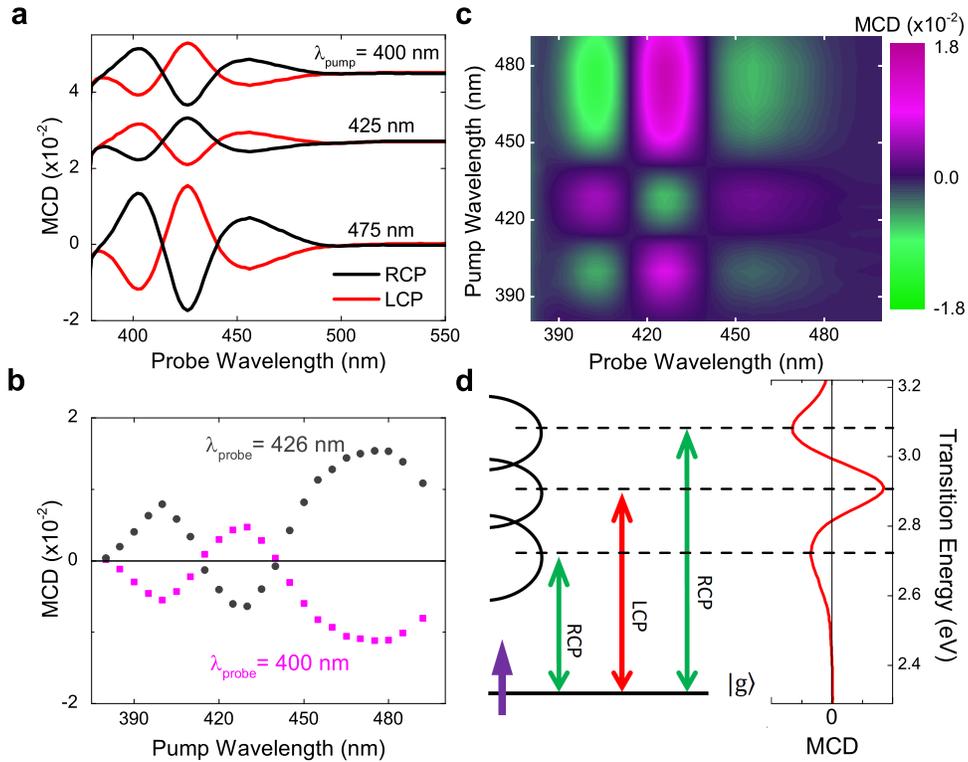}
\caption{\fontsize{10pt}{10pt}\selectfont
\textbf{Controlling the induced magnetization in SrTiO$_{3-\delta}$ by the pump wavelength.~a,} MCD spectra from SrTiO$_{3-\delta}$ ($n=8 \times 10^{17}$~cm$^{-3}$) after being optically pumped at $\lambda_{\rm pump}$=400~nm, 425~nm, and 475~nm ($T$=3~K, $B$=0). The induced magnetization is inverted when $\lambda_{\rm pump}$=425~nm.~\textbf{b,} The induced MCD, detected at $\lambda_{\rm probe}$=400~nm and 426~nm, as a function of $\lambda_{\rm pump}$. The oscillatory behavior mimics that of the measured MCD spectra shown in panel \textbf{a}.~\textbf{c,}~A contour map of the MCD spectra (\emph{x}-axis) at different $\lambda_{\rm pump}$ (\emph{y}-axis).~\textbf{d,}~A possible level diagram, showing a manifold of (at least) three levels optically coupled to a polarizable ground state level $|g\rangle$ (note these levels could also lie below $|g\rangle$). Circularly polarized optical selection rules allow optical pumping and partial orientation of $|g\rangle$. If $|g\rangle$ is oriented spin-down, the selection rules are reversed.  A non-zero MCD signal (right) is produced when these ground states are preferentially polarized.}
\label{Figure3}
\end{figure*}

To demonstrate the potential utility of this persistent magnetization, Fig.~\ref{Figure2}e shows that detailed magnetic patterns can be optically written, stored, and optically read out in SrTiO$_{3-\delta}$.  The acronyms ``LANL" and ``UMN" were written using 400~nm pump light, where the circular polarization (and hence the magnetization direction) was reversed between adjacent letters. Subsequently, the magnetic patterns were read using raster-scanned MCD with $\lambda_{\rm probe}$=425~nm.

Both the magnitude and sign of the optically induced magnetization can also be controlled by the \emph{wavelength} of the pump light, providing insight into the underlying nature of the $V_{\rm O}$-related magnetization.  Figure~\ref{Figure3}a shows MCD spectra acquired after illumination with $\lambda_{\rm pump}$=400, 425, and 475~nm. The signals invert when $\lambda_{\rm pump}$=425~nm, indicating an oppositely oriented magnetization. Fig.~\ref{Figure3}b shows the induced MCD versus $\lambda_{\rm pump}$, measured at $\lambda_{\rm probe}$=400 and 426~nm.  The induced magnetization oscillates with $\lambda_{\rm pump}$, closely tracking the measured MCD spectrum itself (peaks at 400, 430, and 475~nm, and nodes at 415 and 440~nm).  Fig.~\ref{Figure3}c displays how the full MCD spectra evolve as $\lambda_{\rm pump}$ is varied. It is particularly noteworthy that MCD signals at shorter wavelengths are influenced by pump light at longer wavelengths, suggesting a manifold of optical transitions obeying circularly polarized selection rules, which are coupled to a common, optically polarizable ground state as portrayed in Fig.~\ref{Figure3}d.

\begin{figure*} [htbp]
\centering
\includegraphics [scale = 0.70] {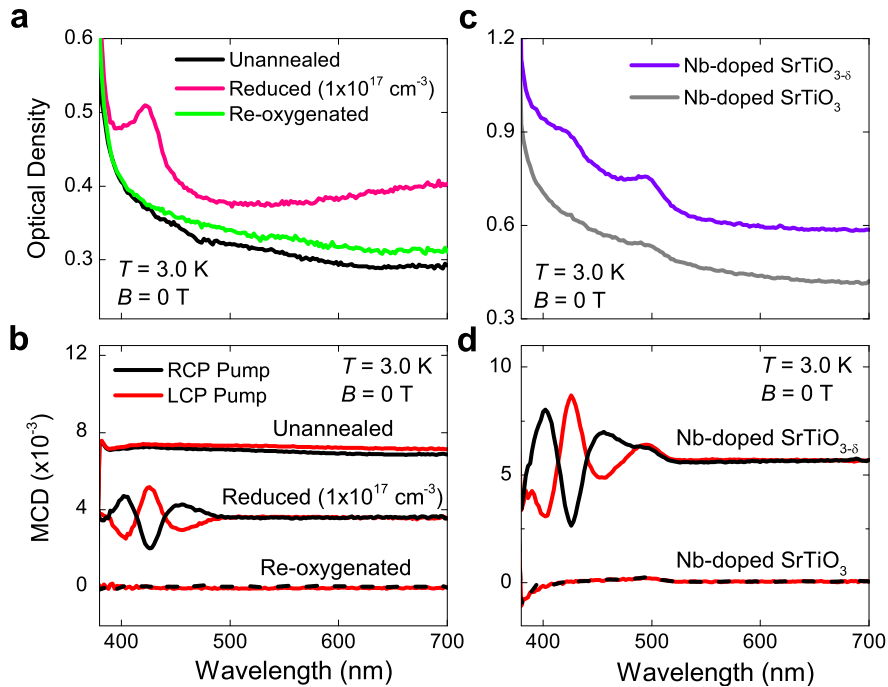}
\caption{\fontsize{10pt}{10pt}\selectfont
\textbf{Magneto-optical properties of SrTiO$_{3}$ and SrTiO$_{3}$:Nb single crystals after creating and removing oxygen vacancies.~a,}~The absorption spectra from an as-received (unannealed) SrTiO$_3$ crystal, again after annealing in UHV to create $V_{\rm O}$, and again after re-oxygenation to remove $V_{\rm O}$. ~\textbf{b,}~The corresponding MCD spectra (offset for clarity) after optically pumping with RCP (black) and LCP (red) light at 400~nm.  Absorption and MCD features disappear after re-oxygenation. \textbf{c,} The absorption of SrTiO$_{3}$:Nb before (grey) and after (purple) introducing $V_{\rm O}$. \textbf{d,} The corresponding MCD spectra (offset for clarity).}
\label{Figure4}
\end{figure*}

To confirm that oxygen vacancies play an essential role in optically induced magnetism, we reduced an as-received (insulating) SrTiO$_3$ substrate and then re-oxygenated it, measuring the optical properties before and after each step (see Figs.~\ref{Figure4}a,b). Only when appreciable $V_{\rm O}$ were present, as determined by a measurable $n$, did we observe sub-bandgap absorption at $\sim$430 nm and optically induced MCD signals. Following re-oxygenation, the absorption disappeared (as also seen in \cite{LeePRB1975}), the crystal was again electrically insulating, and most importantly the MCD signals vanished.

We also investigated the role of excess electrons and concurrent changes in Fermi level by studying SrTiO$_3$ substrates doped with 0.02\% Nb (an electron donor). While SrTiO$_3$:Nb exhibits an absorption feature at 500~nm, optically induced MCD signals were not observed, despite electron densities ($n$$\sim 8\times 10^{17}$ cm$^{-3}$) comparable to our SrTiO$_{3-\delta}$ samples. However, after reducing this substrate (adding $V_{\rm O}$), both a 430~nm absorption feature and optically induced MCD were observed (Figs. 4c,d).  Thus, excess electron density appears to play a minor role in these magneto-optical effects and does not significantly influence magnetization relaxation. This view is further supported by the observation that applied currents (up to 2~mA) have no discernible influence on the optically induced magnetization (not shown). Separately, we checked for surface-related MCD artifacts by mechanically polishing away several microns of both sides of a SrTiO$_{3-\delta}$ sample. The optically induced magnetization was unchanged.

\begin{figure} [htbp]
\centering
\includegraphics [scale = .8] {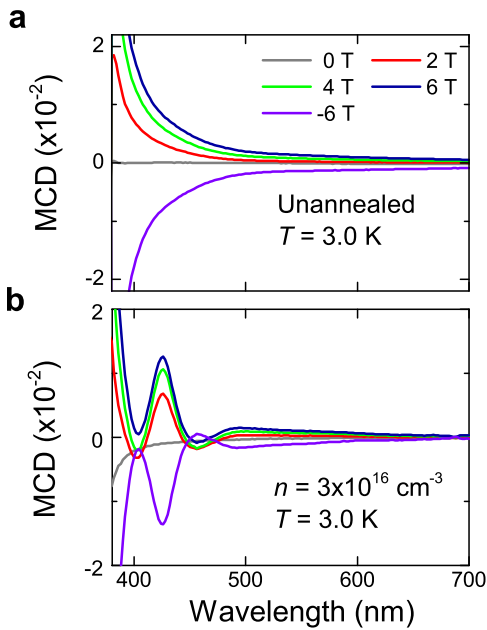}
\caption{\fontsize{10pt}{10pt}\selectfont
\textbf{MCD spectra from SrTiO$_{3-\delta}$ in an applied magnetic field.~a,}~As-received (unannealed) SrTiO$_3$ crystals show only a simple, monotonically decaying sub-bandgap MCD signal when a magnetic field, $B$, is applied. This background MCD grows with $B$ and inverts sign when $B$ is reversed, as expected. $B$ is applied along the sample normal, parallel to the optical axis.~\textbf{b,}~Reduced SrTiO$_{3-\delta}$ samples not only exhibit this background signal, but also show the oscillatory MCD structure that was previously observed under optical pumping.}
\label{Figure5}
\end{figure}

Finally, Fig.~\ref{Figure5} shows MCD spectra at non-zero applied magnetic fields, $B$. As-received SrTiO$_3$ exhibits only a simple, monotonically decaying MCD spectrum below the bandgap, likely arising from band splitting in the absorption tail. This MCD inverts when $B$ is reversed, as expected.  In contrast, reduced SrTiO$_{3-\delta}$ crystals show not only this background, but also the same oscillatory MCD signals that appear after optically pumping at $B$=0~T. Applied magnetic fields therefore polarize these localized complexes in the same manner as optical pumping. Importantly, at fixed $B$ these MCD signals do \emph{not} decay, again consistent with a stable ground state spin polarization.

While these magneto-optical data cannot precisely identify the localized complex responsible for the induced magnetization, some general inferences can be made. All SrTiO$_{3-\delta}$ samples -- independent of total $V_{\rm O}$ density -- exhibit:  i) an additional sub-bandgap optical absorption centered at $\sim$430~nm (Fig.~\ref{Figure1}b), ii) an optically induced MCD having the same spectral shape (Fig.~1c), and iii) the same temperature dependence of the induced MCD (Fig.~1d). These data point to the formation of polarizable $V_{\rm O}$-related states exhibiting sub-bandgap absorption, one candidate being the Fe-$V_{\rm O}$ complexes previously studied in Refs.~\cite{WildPRB1973, LeePRB1975}. Moreover, the extremely long-lived nature of the induced magnetization, its activated relaxation dynamics (Fig.~\ref{Figure2}), and the stability of MCD spectra in non-zero $B$ are consistent with a polarizable ground state. Finally, the oscillatory MCD spectrum, and its dependence on pump wavelength (Fig.~\ref{Figure3}), support a scenario in which this ground state is coupled to a manifold of (at least) three levels split by crystal field and/or Jahn-Teller effects, and spin-orbit coupling. In this picture, circularly polarized optical selection rules apply and some degree of optical orientation of the ground state spin is possible.

As noted above, the polarizable $V_{\rm O}$-related complex may involve other neighboring impurities. Even the highest-quality SrTiO$_3$ crystals currently available contain a variety of impurity atoms, typically with non-negligible concentrations of tens of parts-per-million \cite{SonNatMat2010, LeePRB1975}. Although electron spin resonance studies have established that Fe and other impurities in SrTiO$_3$ can exist in a variety of oxidation states, we do not observe any pump-induced changes in the total optical absorption or any other such `photochromic' behavior \cite{FaughnanPRB1971}. Moreover, the magnetometry data and dependence on circular pump polarization (Fig. 1) rule out effects due to the selective population of defect complexes along different crystal axes (as demonstrated, \emph{e.g.}, for Fe-$V_{\rm O}$ complexes by \emph{linearly} polarized light \cite{BerneyPRB1981}). Rather, our studies represent a fundamentally different phenomenon: the controlled generation of a \emph{net magnetic moment} with circularly polarized light -- in zero magnetic field -- analogous to optical pumping and orientation of electron and hole spins in classic semiconductor spintronic materials like GaAs, but here with extremely long-lived relaxation times.

A cartoon outlining a possible level scheme for this complex is shown in Fig.~\ref{Figure3}d.  Note that the manifold of three levels could also lie below the ground state. (This latter situation would be analogous to strained $n$-type GaAs, where the light-hole, heavy-hole, and split-off valence bands are split by spin-orbit coupling and crystal distortion, so that RCP light pumps electrons with $S_{\rm z} = +\frac{1}{2}, -\frac{1}{2}, +\frac{1}{2}$, respectively.)  Our SrTiO$_{3-\delta}$ MCD data suggest energy splittings of $\sim$200~meV between these optical transitions, which exceeds the splitting recently observed in the $3d$ conduction bands of SrTiO$_3$ surfaces~\cite{SantanderNature2011}, but may be commensurate with $2p$ valence band splitting~\cite{BlazeyPRB1983}. Defect complexes can introduce large local energy scales, particularly if lattice distortions are involved. It is worth noting that local deformations can also play an essential role in long-lived or metastable phenomena, for example persistent photoconductivity from $DX$ centers in III-V semiconductors~\cite{MooneyJAP1990}.

Our data point to exciting possibilities for exploiting magneto-optical effects in perovskite oxides. The ability to optically read, write, and store magnetic information at technologically relevant wavelengths (\emph{e.g.}, 405~nm) in SrTiO$_{3-\delta}$ suggests new opportunities for device applications. This work may also shed new light on possible mechanisms for $V_{\rm O}$-related local moment formation and long-range magnetism at complex oxide interfaces.

\noindent
\newline
\large
\textbf{METHODS}
\newline
\newline
\small
\noindent
\textbf{Oxygen-deficient (reduced) SrTiO$_{3-\delta}$ samples}.
A series of nine 500~$\mu$m thick undoped SrTiO$_3$ (100) crystals from MTI Corp.~were annealed in ultra-high vacuum (oxygen partial pressure $< 10^{-9}$~Torr) at temperatures between 650-750$^{\circ}$C to promote diffusion of oxygen out of the lattice~\cite{SpinelliPRB2010}. Indium contacts were soldered to the corners of each sample in a van der Pauw geometry, and the electron concentration $n$ was measured using longitudinal resistivity and/or Hall studies, from which the approximate $V_{\rm O}$ density was inferred.  For this study, $n$ ranged from $\sim$$3\times 10^{12}$~cm$^{-3}$ to $8 \times 10^{17}$~cm$^{-3}$.  Sub-bandgap absorption and optically induced magnetization were only observed when $n$$>$$10^{14}$ cm$^{-3}$.

\noindent
\textbf{MCD studies of SrTiO$_{3-\delta}$ magnetization.}
MCD spectroscopy was used to obtain a spectrally resolved measure of magnetization in SrTiO$_{3-\delta}$.  MCD detects the normalized difference between the transmission of right- and left-circularly polarized probe light ($T_{\rm R}$ and $T_{\rm L}$, respectively) through the sample:  $\left(T_{\rm R}-T_{\rm L}\right)/\left(T_{\rm R} + T_{\rm L}\right)$.  Non-zero MCD signals generally indicate the presence of time-reversal-symmetry-breaking phenomena (\emph{e.g.}, magnetization). MCD typically arises in zero magnetic field in materials possessing a remnant magnetization (\emph{e.g.}, ferromagnets), or in applied magnetic fields from diamagnetic or paramagnetic materials. A benefit of MCD spectroscopy as compared to global magnetization techniques (such as SQUID magnetometry) is that MCD signals typically occur in specific wavelength ranges, which can help identify the underlying nature of the magnetic species and its coupling to the optical constants of the material. MCD is directly related via Kramers-Kronig relations to the well-known magneto-optical phenomenon of Faraday rotation, which measures magnetic circular birefringence.

The samples were mounted in the variable-temperature (1.5-300~K) insert of an 8~T superconducting magnet with direct optical access. Spectrally narrow, continuous-wave probe light of tunable wavelength was derived from a xenon arc lamp and a 300~mm scanning spectrometer.  The probe light was mechanically chopped, and its polarization was modulated between right- and left-circular by a photoelastic modulator (PEM). Very low optical powers were used (1-100~nW) for the probe. The probe light was weakly focused through the crystals ($\sim$1~mm$^2$ spot area) and was detected by an avalanche photodiode. $T_{\rm R}-T_{\rm L}$ and $T_{\rm R} + T_{\rm L}$ were measured using lock-in amplifiers referenced to the PEM and the chopper, respectively.  Magnetization was induced and controlled in the crystals by a separate, defocused, and independently polarizable pump beam (see Fig.~\ref{Figure1}a).  This light was derived from either a 405~nm laser diode or from a frequency-doubled and wavelength-tunable Ti:sapphire laser. Low pump powers on the order of 5-200~$\mu$W were typically used.  The unfocused pump laser beam globally illuminated the samples with a typical spot area of 13~mm$^2$ for the 405~nm laser diode and 18~mm$^2$ for the doubled Ti:sapphire laser.

\noindent
\textbf{SQUID studies of optically induced magnetization.}
A commercial (Quantum Design MPMS) SQUID magnetometer was used to confirm and quantify the optically induced magnetization. 405~nm light with controlled optical polarization was coupled to the SrTiO$_{3-\delta}$ samples using a single-mode optical fiber. Sample sizes were approximately 3 mm x 3 mm x 0.5 mm.
\newline
\newline
\noindent
\textbf{Acknowledgements}
\newline
We thank D.L.~Smith, Q.~Jia, P. Littlewood and A.V.~Balatsky for helpful discussions.  This work was supported by the Los Alamos LDRD program under the auspices of the US DOE, Office of Basic Energy Sciences, Division of Materials Sciences and Engineering. Work at UMN supported in part by NSF under DMR-0804432 and in part by the MRSEC Program of the NSF under DMR-0819885.


\newpage

\section{Supplemental Information}
\subsection{Temperature Dependent Absorption}
As discussed in the main text, optically induced magnetic circular dichroism (MCD) in SrTiO$_{3-\delta}$ disappeared abruptly when the temperature was raised above $\sim$18~K. To determine if this coincided with any change in the $V_{\rm O}$-related sub-bandgap optical absorption, we measured the optical density of both SrTiO$_{3-\delta}$ and SrTiO$_{3}$ samples over a broad temperature range.

\begin{figure} [htbp]
\includegraphics [scale = .5] {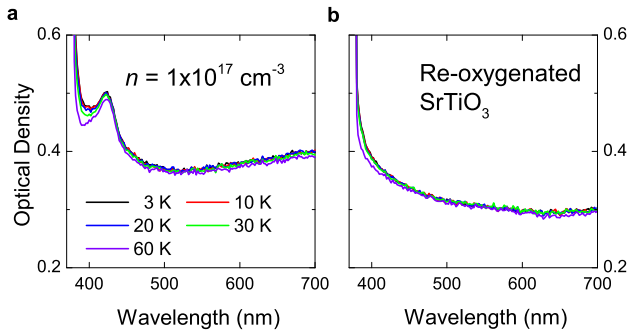}
\caption{\textbf{Supplemental Fig. S1: Temperature dependent optical absorption for SrTiO$_{3-\delta}$ and SrTiO$_3$.} ~The optical density for \textbf{a}, oxygen-deficient SrTiO$_{3-\delta}$ and \textbf{b}, re-oxygenated SrTiO$_3$ changes only minimally as the temperature is increased from 3~K to 60~K. In particular, the $V_{\rm O}$-related sub-bandgap absorption at $\sim$430~nm remains largely unchanged, despite the disappearance of magneto-optical phenomena at temperatures exceeding 18~K.}
\label{SupplFigure1}
\end{figure}

Figures \ref{SupplFigure1}a and \ref{SupplFigure1}b show the sub-bandgap absorption spectra for a SrTiO$_{3-\delta}$ sample and for the same sample after it has been re-oxygenated.  For both samples, the optical density is only minimally altered as the temperature is raised from 3 to 60~K; the sub-bandgap absorption features that are related to oxygen vacancies do not vanish above 18~K.

\subsection{SQUID studies of optically induced magnetization}

\begin{figure} [htbp]
\includegraphics [scale = .5] {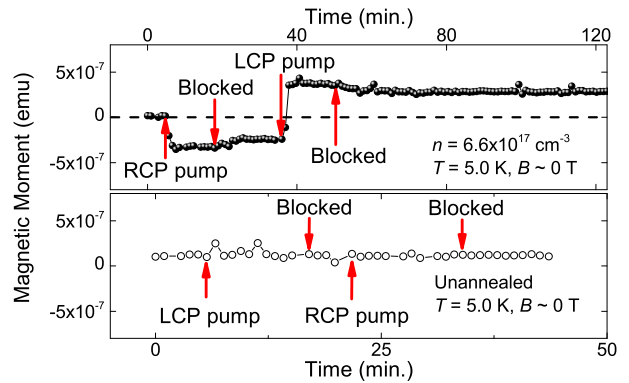}
\caption{\textbf{Supplemental Fig. S2: SQUID measurements of the optically induced magnetic moment in SrTiO$_{3-\delta}$ and in as-received (unannealed) SrTiO$_3$.} Pump polarization dependence of the magnetization of SrTiO$_{3-\delta}$ (upper panel) and SrTiO$_3$ (lower panel) shows that only SrTiO$_{3-\delta}$ exhibits optically induced magnetization.  In a similar manner to the MCD results, the optically induced magnetization in SrTiO$_{3-\delta}$ persists long after the pump illumination is blocked.}
\label{SupplFig3}
\end{figure}

Optically induced magnetization in SrTiO$_{3-\delta}$ was measured using a commercial SQUID magnetometer (Quantum Design MPMS) in zero magnetic field.  Polarization-controlled 405 nm pump light was coupled to the sample via single-mode optical fiber. The end of the fiber was positioned several centimeters above the sample, so that the pump light globally illuminated the sample. The top panel of Fig.~\ref{SupplFig3} demonstrates the ability to manipulate the magnetization of the sample with polarized light, just as the MCD results showed in Fig.~2a of the main text.  As a control, we also tested an unannealed, as-received SrTiO$_3$ sample.  No optically induced magnetization was observed, in agreement with the MCD results presented in Fig.~1c of the main text.  The small magnetization offset in the data is attributed to the diamagnetic response of SrTiO$_3$ created by the small remnant field ($<$10$^{-4}$~T) present in the superconducting magnet of the SQUID magnetometer.

\subsection{Pump Power Dependent Measurements}
\begin{figure} [htbp]
\includegraphics [scale = .5] {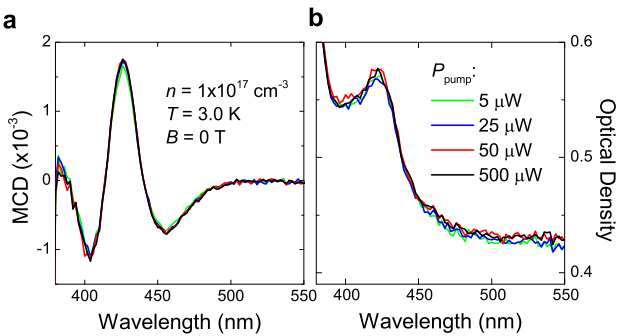}
\caption{\textbf{Supplemental Fig. S3: Saturation of MCD spectra at different pump intensities.~a,} MCD spectra at $B$=0~T measured after optical pumping with 405 nm light of different power (5, 25, 50, and 500~$\mu$W). The spectra are approximately identical, indicating a saturation of the optically induced magnetization.  (From each of these spectra the small, but non-zero, MCD spectrum acquired without optical pumping was subtracted.) \textbf{b,}~The corresponding optical absorption spectra.}
\label{SupplFig4}
\end{figure}

Fig.~2b of the main text shows that the equilibration rate of optically induced magnetization depends strongly on the intensity of the pump illumination, and that the induced magnetization saturates at approximately the same magnitude independent of pump intensity.  This behavior is consistent with a material containing a fixed density of ground state levels in the gap that are optically polarizable and that have extremely long spin relaxation times.  Fig.~\ref{SupplFig4}a shows that the MCD signals saturate not just at one wavelength, but over the entire MCD spectrum, independent of pump intensity.  These data further support the scenario described in the main text: a manifold of (at least) three circularly polarized optical transitions are coupled to a common ground state, and the buildup of polarization in this ground state affects all three optical transitions simultaneously.

As discussed in the main text, even though the magnetization (MCD) is saturated, the MCD signals themselves are small: the differential change in transmission between RCP and LCP light is only of order $10^{-3}$. There is very little change in the \emph{total} optical absorption in the 400-500~nm range, as shown in Fig.~\ref{SupplFig4}b.  This observation is consistent with the absence of long-lived \emph{excited} states, whose presence would likely create more significant changes in the overall optical density.


\subsection{Slow Temporal Dynamics of Optically Induced Magnetism}

\begin{figure} [htbp]
\includegraphics [scale = .5] {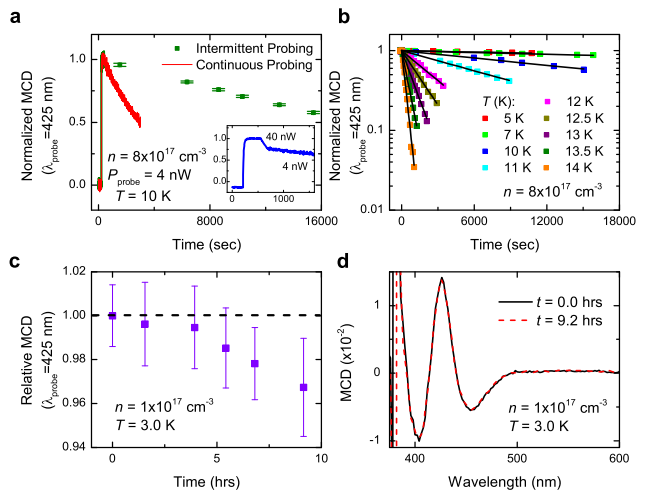}
\caption{\textbf{Supplemental Fig. S4: Temporal dynamics of magnetization relaxation in SrTiO$_{3-\delta}$.~a,}~Continuous probing (red trace) and intermittent probing (green points) of the MCD signal after optical pumping with RCP light ($\lambda_{\rm pump}$=405~nm, $\lambda_{\rm probe}$=425~nm).  The faster decay with continuous probing shows that the MCD measurement itself can accelerate the relaxation of the optically induced magnetization.  Inset: MCD signal as a function of time after optical pumping, using 40~nW and 4~nW of probe power. The higher probe power clearly produces a faster decay of the magnetization.~\textbf{b,}~Intermittent probing of the 425 nm MCD signal after 405 nm optical excitation, for several temperatures.  The black lines indicate single-exponential fits to the data.~\textbf{c,}~Measuring the relaxation of optically induced magnetization at 3~K over the course of $\approx$9 hours. Only a small decay of approximately 4\% is observed over that time span.  A probe power of 0.4~nW was used in order to minimize probe-induced relaxation.~\textbf{d,}~The full MCD spectrum measured just after magnetizing the sample (black line) and 9.2 hours later (red dashed line) shows that the induced magnetization at 3 K has barely changed.  As with \textbf{c}, 0.4~nW of probe power was used to minimize probe-induced relaxation.}
\label{SupplFig2}
\end{figure}

The extremely slow relaxation of optically induced magnetism in SrTiO$_{3-\delta}$ suggests a potential for magneto-optical information storage at low temperatures.  However, the ability to induce and control magnetization with polarized pump light also means that optical probes of magnetization (such as MCD or Faraday rotation) can potentially perturb the magnetization. Thus, it is important to recognize and to minimize any influence of the \emph{probe} light on the magnetization when measuring the relaxation dynamics of SrTiO$_{3-\delta}$. One way to accomplish this is to use very low intensities of probe light. As Fig.~\ref{SupplFig2}a shows, the act of \emph{continuously} probing the MCD (here using $\lambda_{\rm probe}$=425~nm) causes an optically induced magnetization to relax more quickly than if probed intermittently.  Unsurprisingly, the relaxation rate is faster when using higher probe intensity, as the inset displays. Therefore, in both Figs.~\ref{SupplFig2}a and b, we used a very weak ($P_{\rm probe}$ = 4 nW) probe beam to measure the magnetization at discrete intervals.  Each point shown in Figs.~\ref{SupplFig2}a-c is an average of data collected over tens of seconds with the error bars denoting the standard deviation (error bars are not shown in Fig.~\ref{SupplFig2}b for clarity).  We investigated temperatures from 20 to 3~K, with 17.5~K being the fastest magnetization decay that we could reliably measure.  From 3 to 14~K, probing was performed intermittently, while above those temperatures (as given in Fig.~2c in the main text) a continuous probe was employed.

Whether continuously or intermittently probed, each data trace was normalized to unity during the optically pumped portion of the curve (the first 100 seconds).  A single exponential decay, $\exp\left(-t/\tau\right)$, fit every data trace quite well, allowing us to extract $\tau$ as a function of temperature (these values of $\tau$ are shown in Fig.~2d of the main text).  $\tau$ approximately follows an activated (Arrhenius) behavior for temperatures between 13~K and 17~K, but deviates significantly from activated behavior below $\sim$10~K.  We note however that owing to the extremely long magnetization relaxation times in this low-temperature regime, even very tiny amounts of inadvertent light leaking on to the samples could account for this behavior.

In a different set of experiments, a SrTiO$_{3-\delta}$ sample ($n=1\times10^{17}$~cm$^{-3}$) was magnetized in a 6~T magnetic field. The field was then ramped to zero, and a MCD spectrum was then acquired using a 0.4~nW probe.  Over the next $\approx$9~hours, the sample magnetization was intermittently probed by MCD (at 425~nm) for tens of seconds at a time, after which another full MCD spectrum was acquired. Figure~\ref{SupplFig2}c shows the averaged results of the intermittent probing, in units of relative decay from a starting normalized value of 1.  Remarkably, the MCD decays by less than 4\% over the course of 9 hours, demonstrating the persistence of the induced magnetization at $B=0$~T.  A comparison of the full MCD spectra at $t$ = 0 and at $t$ = 9.2 hours is given in Fig.~\ref{SupplFig2}d; the spectra are essentially identical. This ability to induce a long-lived magnetization using either circularly polarized light \emph{or} a magnetic field is consistent with a scenario in which magnetization in SrTiO$_{3-\delta}$ originates from polarizable ground state levels with an extremely long relaxation time.

\subsection{Photoluminescence Measurements}

\begin{figure} [htbp]
\includegraphics [scale = 0.5] {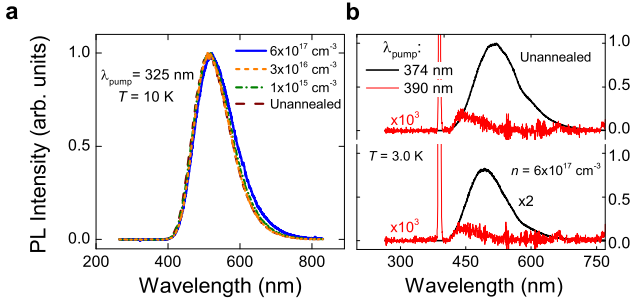} 
\caption{\textbf{Supplemental Fig. S5: Photoluminescence (PL) from SrTiO$_{3-\delta}$ using above- and below-bandgap excitation.~a,}~Using weak above-bandgap excitation (325~nm), normalized PL spectra from substrates with different $V_{\rm O}$ densities shows that the PL remains largely unaffected by oxygen removal from the lattice.~\textbf{b,}~Comparison of PL from an unannealed, as-received sample and an oxygen-deficient sample, for weak excitation both above-bandgap (374 nm) and below-bandgap (390 nm). Below-bandgap excitation generates negligible PL relative to above-bandgap excitation.  Regardless of excitation wavelength, the PL from unannealed and from reduced SrTiO$_3$ looks qualitatively similar.}
\label{SupplFig5}
\end{figure}

Photoluminescence (PL) was excited using either a He-Cd laser (325~nm) or a frequency-doubled Ti:sapphire laser (tunable wavelengths from 360-490~nm). The latter permits either above-bandgap or below-bandgap excitation. A series of oxygen-deficient SrTiO$_{3-\delta}$ samples and an unannealed (as-received) SrTiO$_{3}$ sample was measured. For the case of 325 nm (above-gap) excitation, all samples showed a very broad and very similar PL band peaked well below the band-edge (see Fig.~\ref{SupplFig5}a), consistent with prior results and likely due to self-trapped excitons or defects. No significant differences between oxygen-deficient and as-received samples was observed in the PL.

Figure~\ref{SupplFig5}b shows the marked difference in PL between above- and below-bandgap excitation, for both as-received SrTiO$_3$ (top) and for SrTiO$_{3-\delta}$ (bottom).  Here, the PL intensities for each sample were scaled by the integration time and then normalized to the PL peak of the unannealed SrTiO$_3$ substrate.  For both samples, the below-bandgap PL is negligible when compared to above-bandgap excitation, and no significant differences between the samples are observed.

\end{document}